%% file: main.tex
\documentclass[journal]{IEEEtran}

\usepackage{amsmath,amssymb}
\usepackage{graphicx}
\usepackage{booktabs}
\usepackage{cite}
\usepackage{url}

\title{Multidimensional Source-Space MEG
Representation of Network Imbalance in ADHD}

\author{
Xiran~Guo, Zhihao~Liu, and Guanghua~Liu,~\IEEEmembership{Senior Member,~IEEE}%
\thanks{Xiran Guo, Zhihao Liu, and Guanghua Liu are with the School of Cyber Science and Engineering, Huazhong University of Science and Technology, Wuhan 430074, China (e-mail: guoxiran@hust.edu.cn; d202480852@hust.edu.cn; guanghualiu@hust.edu.cn).}%
\thanks{Corresponding author: Guanghua Liu.}%
}

\begin{document}

\maketitle


\noindent{\footnotesize\itshape
This work has been submitted to the IEEE for possible publication.
Copyright may be transferred without notice, after which this
version may no longer be accessible.\par}

\vspace{0.5em}


\begin{abstract}
Attention-deficit/hyperactivity disorder (ADHD) is associated with an atypical
imbalance between the default mode network (DMN) and task-control systems. However,
compressing neural activity into a single macroscopic imbalance measure may
obscure individual neurophysiological differences. To address this, we
leveraged the source-orientation selectivity and related measurement properties of
resting-state magnetoencephalography (MEG) to establish a multidimensional source-space
network representation in 21 individuals with ADHD and 21 healthy controls. We defined
the macroscopic imbalance in relative Alpha activity between the DMN and task-control
systems as the Core, and expanded its constituent information along three
dimensions—spatial sources, spectral composition, and temporal network
organization—into in-depth measures, assessed for stability, robustness, and
complementarity. We found that the ADHD group exhibited a significant Core Alpha
imbalance, driven mainly by task-control regions, while the multidimensional
expansion revealed complementary ADHD-related neurophysiological information not
fully represented by the Core, suggesting that similar macroscopic imbalances
may reflect distinct underlying physiological states. Across 40 exploratory
within-cohort held-out evaluations with separate development and test subsets,
mean ROC-AUC increased from 0.7607 (Core-only) to 0.8995 with multidimensional
inputs and to 0.9051 with the final ensemble, and mean accuracy increased from
71.43\% to 79.82\%. These findings suggest that multidimensional source-space
expansion can further clarify the neurophysiological composition of ADHD-related
network imbalance and improve individual-level discrimination, offering an
interpretable framework potentially applicable to other brain disorders
characterized by network imbalance.
\end{abstract}



\input{introduction}
\input{methods}
\input{results}
\input{discussion}
\input{conclusion}


\section*{Data Availability Statement}
The data were obtained from the Open MEG Archive
(OMEGA), subject to its registration and data-use
conditions; see
\url{https://omega.bic.mni.mcgill.ca}
and~\cite{ref32}.


\bibliographystyle{IEEEtran}
\bibliography{references}

\end{document}

%% file: introduction.tex
\section{Introduction}

Attention-deficit/hyperactivity disorder (ADHD)
is a common neurodevelopmental disorder affecting
approximately 7\% of children worldwide~\cite{ref01,ref02}.
It is characterized by persistent inattention
and/or hyperactivity-impulsivity, with substantial
individual variability~\cite{ref01}.
ADHD identification still relies primarily on
clinical interviews, behavioral observations,
and standardized rating scales.
Although essential for clinical diagnosis,
these approaches mainly capture behavioral
manifestations and provide limited insight into
underlying neurophysiological mechanisms.
Identifying stable, reproducible, and
physiologically interpretable brain measures
to support more objective individual-level
assessment therefore remains an important
goal in ADHD research~\cite{ref01}.

Among the functional abnormalities reported
in ADHD, imbalance between the default mode
network (DMN) and task-control-related networks
is particularly relevant.
The DMN primarily supports internally directed
processes, whereas task-control networks support
externally directed attention and cognitive
control~\cite{ref03,ref04}.
Their coordination is essential for maintaining
and shifting attentional states.
Previous fMRI and electrophysiological studies
have repeatedly identified abnormal activity
and connectivity within the DMN and
task-related networks in ADHD~\cite{
ref05,ref06,ref07,ref08,ref09,ref10,ref11,ref12,
jbhi2024_spatial,jbhi2026_dynamic}.
These abnormalities have also been linked to
genetic variation and treatment response~\cite{
ref13,ref14,ref15}, while related network features
have been used for individual-level ADHD
identification~\cite{
ref16,jbhi2022_threshold,jbhi2025_fusion,
jbhi2026_spatiotemporal}.
Thus, DMN--task-control imbalance provides
a low-dimensional summary of complex network
abnormalities, potentially linking altered
brain function to individual-level diagnosis.


However, this low-dimensional summary creates
an information bottleneck.
Similar imbalance values may occur in both
ADHD and control individuals: some with ADHD
show little imbalance, whereas some controls
exhibit comparable patterns~\cite{ref17,ref18}.
As later illustrated in Fig.~2A, although the
DMN--task-control Alpha imbalance measure
(hereafter, Core) shows a group-level shift,
substantial overlap remains between ADHD
and control individuals at the subject level.
This overlap suggests that compressing complex
network states into a single imbalance measure
may obscure neurophysiological differences
between individuals, mapping distinct
physiological states onto similar Core values
and limiting individual-level discrimination.
The next step is therefore to analyze
the neural activity that gives rise to
the observed imbalance, rather than relying
on the difference between the two networks alone.


Such an analysis requires a measurement
approach that can resolve the spatial
sources of neural activity while preserving
its rapid spectral and temporal dynamics.
Among fMRI, EEG, and MEG, which are
commonly used to investigate brain
networks, MEG offers measurement
properties particularly suited to
this multidimensional analysis.
Unlike the slow blood-oxygen-level-dependent
response of fMRI, MEG measures magnetic
fields generated by neural currents,
providing direct access to rapid
electrophysiological activity~\cite{ref19,ref20,ref21,ref22}.
Compared with EEG, MEG is generally less
sensitive to skull conductivity uncertainties
in forward modeling and source
estimation~\cite{ref23,ref24,ref25}
and exhibits distinct source-orientation
selectivity~\cite{ref26}, supporting
cross-subject source-space analysis~\cite{jbhi2025_source}.
Together, these properties provide a
measurement foundation for resolving
the spatial contributions of the Core
and characterizing complementary spectral
and temporal-network information.
We further evaluated these measurement
properties through matched EEG--MEG
simulations to assess their contributions
to source-space network recovery.


Building on the source-space measurement foundation provided by MEG, we addressed the information bottleneck of the Core by extracting complementary neurophysiological information from the underlying neural signals. Specifically, we expanded the Core representation along three physiologically connected dimensions.
First, similar Core values may arise from
different DMN and task-control contributions,
requiring spatial decomposition to resolve
this ambiguity.
Second, because ADHD-related abnormalities
extend beyond a single frequency
band~\cite{ref10,ref27}, the Core cannot fully
capture their spectral characteristics.
Different oscillatory rhythms are associated
with distinct neural processes, including
Alpha-related information gating and
Theta-related cognitive control, while
aperiodic measures characterize broadband
spectral properties~\cite{ref28,ref29,ref30,ref31}.
We therefore combined these measures to
examine complementary neurophysiological
information beyond the Core.
Third, time-averaged spectra cannot capture
network reconfiguration or stability, whereas
MEG preserves the temporal information needed
to characterize dynamic network coordination.
These spatial, spectral, and temporal measures
were linked by their source-space networks
but captured distinct physiological properties.
Core-adjusted analyses then tested whether
the expanded measures provided additional
ADHD-related information rather than merely
restating the Alpha contrast.


Overall, this study established a source-space
MEG framework that resolves the spatial sources
of network imbalance, incorporates complementary
spectral and temporal information, and assesses
its value for individual-level ADHD discrimination.
At the measurement level, we examined MEG
properties relevant to rapid neural oscillations
and cross-subject source reconstruction.
At the representation level,
DMN--task-control Alpha imbalance was expanded
along spatial-source, spectral, and
temporal-network dimensions.
At the validation level, temporal reliability,
technical-factor adjustment, empirical
sensor-noise perturbation, and Core-adjusted
complementarity analyses assessed the stability,
robustness, and complementarity of the
expanded information.
Finally, we evaluated whether the
multidimensional representation improved
individual-level discrimination through
repeated stratified train--test splits,
with model development confined to each
training subset and diagnostic performance
summarized across held-out evaluations,
emphasizing ROC-AUC as the principal measure
of discriminative performance.

%% file: methods.tex
\section{Materials and Methods}

Resting-state MEG signals were first reconstructed
into individual source spaces to obtain
region-of-interest (ROI) cortical time series
and derive the core Alpha imbalance between
the default mode network (DMN) and
task-control-related regions (Core).
Complementary information was then extracted
along spatial-source, spectral, and
temporal-network dimensions and evaluated
for temporal reliability, technical influences,
sensor-noise robustness, and Beyond-Core
complementarity.
Finally, individual-level ADHD discrimination
was assessed using repeated stratified
train--test splits, with a separately held-out
test subset in each split.

\subsection{Participants and Data Acquisition}

This study retrospectively analyzed data from
the Open MEG Archive (OMEGA)~\cite{ref32}.
The final sample comprised 42 participants
(21 ADHD and 21 healthy controls), classified
according to OMEGA diagnostic labels.
Only participants with complete resting-state
MEG, structural MRI, and spatial information
required for source reconstruction were included.
Selected resting-state recordings were used
for the primary analyses.

\noindent\textit{Ethics and Data Use Statement:}
This study is a retrospective secondary analysis
of previously collected, de-identified
resting-state MEG data from the Open MEG
Archive (OMEGA)~\cite{ref32}, accessed under
an approved request and the OMEGA Data Use
Agreement. All OMEGA participants were recruited
for other MEG research studies and were offered
to share their resting-state MEG data in OMEGA,
and they signed the OMEGA consent form.
The McGill University Health Centre Research
Ethics Board reviewed the OMEGA registry
and is responsible for its ethical oversight;
under the OMEGA Data Use Agreement, no
additional authorization was required for
this secondary analysis. No new participants
were recruited or contacted, and no additional
human-subject measurements were performed.
In accordance with the Data Use Agreement,
only group-level results are reported; no
single-participant data and no participant
ID codes are presented. The data were kept
in a secure environment and were not
redistributed; requests for access should
be directed to OMEGA.

Mean age was $21.75\pm4.72$ years in ADHD and
$23.62\pm6.25$ years in controls; the groups comprised 8 males/13 females
and 12 males/9 females, respectively
(Fisher's exact $p=0.3543$). Age was obtained
for the MEG session analyzed. Demographic sensitivity
analyses assessed age and sex as covariates.

MEG data were acquired in a magnetically
shielded room using a whole-head 275-channel
CTF system at 2400 Hz, with a 600-Hz hardware
anti-aliasing low-pass filter.
ECG and horizontal and vertical EOG were
recorded simultaneously.
Head shape and anatomical landmarks were
digitized using a Polhemus Fastrak system
and registered to individual T1-weighted MRI.

\subsection{MEG Preprocessing and
Source-Space Reconstruction}

\subsubsection{MEG Preprocessing}

The 265 MEG channels common to all participants
were retained.
Preprocessing comprised 1--40 Hz band-limiting,
50-Hz line-noise suppression,
and resampling to 200 Hz.
From each participant, 250 s of continuous
resting-state data were extracted;
after 5-s edge trimming, the remaining
240 s were divided into eight nonoverlapping
30-s windows.
Theta and Alpha bands were defined as
4--8 Hz and 8--12 Hz, respectively.
These windows were used for window-based
analyses, whereas Theta network-role
stability was derived separately from
the first 30 s of each participant's data.

\subsubsection{Source Reconstruction and
ROI Time Series}

Individual T1-weighted MRI scans were
processed with FreeSurfer for cortical
surface reconstruction and aparc anatomical
parcellation~\cite{ref33}.
After MEG--MRI coregistration, individual
cortical source spaces and forward models
were constructed.
Noise covariance was estimated from the
corresponding noise recordings, and source
activity was reconstructed using dynamic
statistical parametric mapping (dSPM)
within the minimum-norm framework.
The inverse operator used a loose-orientation
constraint of 0.2, depth weighting of 0.8,
and $\lambda^2=1/9$, retaining cortical
surface-normal components
(\texttt{pick\_ori="normal"}).
These parameters were fixed before analysis
and were not tuned using group differences
or classification performance.

For each aparc ROI, vertex-level dSPM
time series were combined using the
MNE-Python \texttt{mean\_flip} method,
which corrects signal polarity based
on source orientation before averaging
to reduce cancellation caused by
cortical folding.
ROI aggregation preceded all band-limited
and spectral analyses.

\subsection{Leveraging MEG Measurement
Properties for Source-Space Representation}

Matched EEG--MEG models were used to assess
how conductivity uncertainty and source
orientation affect source-space network recovery.
Ground-truth 10-Hz source activity was generated,
with DMN and task-control-related amplitudes
varied to simulate different network states.
Identical source states were projected through
EEG and MEG forward models and reconstructed
using the dSPM settings adopted in the
main analysis.
Simulations lasted 2 s at 200 Hz.

Conductivity analysis involved six participants
with complete individual models.
Reference skull conductivity was 0.006 S/m,
with perturbations of 0.003 and 0.012 S/m.
Sensor data generated under the reference
model were reconstructed using the reference
and both perturbed models.
Recovery errors for task-control activity,
DMN activity, and Core were measured by
normalized mean absolute error (nMAE),
using the across-state standard deviation
of the reference representation for
normalization.

Source-orientation analysis also involved
six participants.
At nine representative cortical locations,
source position and strength were fixed.
Local radial directions were defined relative
to a reference center fitted to the individual
MRI surface, together with two orthogonal
tangential directions.
EEG and MEG responses were evaluated for all
three orientations to derive normalized radial
response, tangential preference, and
direction selectivity.
The reference center was additionally perturbed
to assess the stability of orientation measures.
Radial and tangential basis vectors were
constructed orthogonally.

\subsection{Core DMN--Task-Control Alpha Imbalance}

ROIs were predefined using the Desikan--Killiany parcellation and previous DMN and task-control network studies~\cite{ref04,ref33,ref34,ref35,ref36}. As the resting-state dataset lacked individual functional localizers, these anatomical ROIs approximated the corresponding functional systems. DMN ROIs comprised the bilateral precuneus, posterior cingulate, and medial orbitofrontal cortex. The frontal-control component of the Core comprised the bilateral rostral middle frontal, caudal middle frontal, and superior frontal cortices. Bilateral superior and inferior parietal cortices formed a separate dorsal-attention ROI set and were not included in the Core.

Alpha activity was selected to characterize relative network organization because of its associations with information gating and functional inhibition~\cite{ref30}. The Core was designed as a low-dimensional summary of relative Alpha amplitudes rather than an exclusive measure of network imbalance.

ROI-level dSPM time series were filtered to 8--12 Hz. For each 30-s window $w$, Alpha activity in ROI $r$ was quantified by root-mean-square (RMS) amplitude:

\begin{equation}
R_{r,w}^{\alpha} =
\sqrt{\frac{1}{T_w}\sum_{t\in w}
\left[x_r^{\alpha}(t)\right]^2}.
\label{eq:alpha_rms}
\end{equation}

For each system $S$, ROI log-RMS amplitudes were averaged within each window:
\begin{equation}
A_{S,w}^{\alpha} =
\frac{1}{N_S}\sum_{r\in S}
\ln(R_{r,w}^{\alpha}+\varepsilon).
\end{equation}
Here $N_S$ is the number of ROIs in system $S$. These measures are noise-normalized dSPM log amplitudes, not absolute currents in nAm. The window-level Core was defined as:
\begin{equation}
C_w = A_{\mathrm{Control},w}^{\alpha}
-A_{\mathrm{DMN},w}^{\alpha}.
\label{eq:core}
\end{equation}
Here, Control specifically denotes the frontal-control ROI set. This difference reduces sensitivity to common multiplicative scaling. Subject-level Core was the median across eight windows. The two network log-Alpha components were retained separately because distinct component values can yield similar Core values.

\subsection{Multidimensional Expansion of the Core Imbalance}

\subsubsection{Spatial-Source Dimension}

The frontal-control and DMN Alpha components of the Core were retained separately. Because the Core is a difference, resolving its two terms distinguishes the frontal-control and DMN contributions underlying similar imbalance values. Both were calculated using the same 30-s windows and subject-level median aggregation as the Core, followed by between-group comparisons.

\subsubsection{Spectral Dimension}

Frontal-control Theta was defined over 4--8 Hz. Source time series from predefined frontal-control ROIs were bandpass-filtered, and ROI-level RMS amplitudes were log-transformed and averaged within the network in each window. The subject-level measure was the median across eight windows. The Theta band was selected based on previous cognitive-control research~\cite{ref31}. Frontal-control Theta therefore probes control-related oscillatory activity outside the Alpha band summarized by the Core.

Aperiodic spectral features were estimated from source-space power spectral density (PSD) using Welch's method with Hann windows, 4-s segments, and 2-s overlap. For the inter-network gradient, a log--log linear fit was performed over 2--7 and 14--30 Hz to exclude the main Alpha peak:

\begin{equation}
\log_{10}P(f)=a+b\log_{10}f,\quad E=-b.
\label{eq:aperiodic}
\end{equation}

The DMN and frontal-control regions comprised the bilateral precuneus and rostral middle frontal cortex, respectively. The window-level aperiodic gradient was defined as:

\begin{equation}
G_{\mathrm{ap},w}
=E_{\mathrm{Frontal},w}-E_{\mathrm{DMN},w}.
\label{eq:aperiodic_gradient}
\end{equation}

Subject-level gradients were the medians across eight windows. The superior-parietal exponent was estimated separately over 2--30 Hz, excluding 4--13 Hz to reduce Theta- and Alpha-band contributions. Because the gradient fitting range retains 4--7 Hz activity, these exponents were interpreted as empirical spectral-slope measures rather than fully isolated aperiodic components. The gradient probes spectral differences within the two Core systems, whereas the superior-parietal exponent extends spectral assessment to attention-related cortex outside the Core.

\subsubsection{Temporal-Network Dimension}

The Theta network comprised four predefined nodes: bilateral precuneus and rostral middle frontal ROIs, without data-driven node selection. ROI signals were filtered to 4--8 Hz, and pairwise phase-locking values (PLVs) were calculated from Hilbert phases to construct weighted connectivity matrices. Weighted participation coefficients were then calculated using the predefined DMN and frontal-control modules.

For each participant, the first 30 s of data were divided into two 15-s segments, yielding node participation coefficient vectors $\mathbf{PC}^{(1)}$ and $\mathbf{PC}^{(2)}$. Theta-band network-role stability was defined as:

\begin{equation}
S_{\mathrm{role}}=
\frac{1}{
1+\operatorname{mean}\left(
\left|\mathbf{PC}^{(1)}
-\mathbf{PC}^{(2)}\right|
\right)}.
\label{eq:role_stability}
\end{equation}

Higher values indicate greater consistency of node network roles between the two segments. Unlike the time-averaged Alpha and spectral measures, role stability characterizes short-term consistency of network participation. One stability value was obtained per participant.

The seven measures (Table~\ref{tab:source_measures}) are network-anchored: spatial terms define the window-level Core, while spectral and temporal measures probe the same DMN--frontal systems or related parietal cortex through additional properties. Classification used predefined combinations without assuming equal importance.

\begin{table*}[!t]
\centering
\caption{Multidimensional Source-Space Measures and Their Physiological Rationale}
\label{tab:source_measures}
\footnotesize
\renewcommand{\arraystretch}{1.18}
\setlength{\tabcolsep}{4pt}

\begin{tabular}{
p{0.085\textwidth}
p{0.205\textwidth}
p{0.185\textwidth}
p{0.285\textwidth}
p{0.16\textwidth}
}
\toprule
\textbf{Dimension} &
\textbf{Measure} &
\textbf{ROIs} &
\textbf{Computation} &
\textbf{Physiological Rationale} \\
\midrule

Core &
Core Alpha imbalance &
DMN and frontal-control &
8--12 Hz ROI log-RMS mean; network difference; eight-window median &
Low-dimensional Alpha contrast between networks \\

Spatial &
Frontal-control Alpha &
Frontal-control ROIs &
8--12 Hz ROI log-RMS mean; eight-window median &
Frontal contribution underlying the Core \\

Spatial &
DMN Alpha &
DMN ROIs &
8--12 Hz ROI log-RMS mean; eight-window median &
DMN contribution underlying the Core \\

Spectral &
Frontal-control Theta &
Frontal-control ROIs &
4--8 Hz ROI log-RMS mean; eight-window median &
Control-related oscillations beyond Alpha \\

Spectral &
Inter-network aperiodic gradient &
Bilateral precuneus and rostral middle frontal &
2--7 and 14--30 Hz log--log spectral fitting; frontal minus DMN exponent; eight-window median &
Inter-network empirical spectral-slope differences \\

Spectral &
Superior-parietal aperiodic exponent &
Superior parietal cortex &
Log--log PSD fitting over 2--30 Hz, excluding 4--13 Hz &
Attention-related regional spectrum beyond the Core \\

Temporal &
Theta network-role stability &
Bilateral precuneus and rostral middle frontal &
4--8 Hz PLV; weighted participation coefficients; first 30 s split into two 15-s segments &
Short-term consistency of network participation \\

\bottomrule
\end{tabular}

\vspace{3pt}
\parbox{0.95\textwidth}{
\footnotesize
DMN: default mode network;
ROI: region of interest;
RMS: root-mean-square;
PSD: power spectral density;
PLV: phase-locking value.
The seven measures characterize the Core and its multidimensional expansion. Candidate feature subsets were specified for exploratory comparison; fusion weights and decision thresholds were selected within each development subset.
}
\end{table*}

\subsection{Reliability, Robustness, and Complementarity Analyses}

\subsubsection{Temporal Reliability}

Temporal reliability was evaluated across eight nonoverlapping 30-s windows using ICC(2,1) and inter-window consistency. Participants, rather than individual windows, were treated as independent units in all group-level analyses.

\subsubsection{Technical-Factor Adjustment}

Technical covariates were selected from MEG--MRI coregistration distance, missing-channel count, available data-channel count, and total MEG-channel count. To reduce collinearity, candidate variables with absolute pairwise Spearman correlation $\geq 0.90$ with an already retained variable were excluded. Coregistration distance (mm) and missing MEG-channel count were retained.

For each major measure, the following model was fitted:

\begin{equation}
\begin{aligned}
X_i ={}& \beta_0+\beta_1D_i+\beta_2M_i\\
      &+\beta_3R_i+\varepsilon_i.
\end{aligned}
\label{eq:technical_adjustment}
\end{equation}

Here, $D_i$ denotes diagnostic group, $M_i$ the number of missing MEG channels, and $R_i$ the MEG--MRI coregistration distance in millimeters. For the technical-factor regression, outcomes and continuous covariates were standardized, and ordinary least squares coefficients were tested using conventional residual-variance standard errors. The adjusted diagnostic-group coefficient $\beta_1$ was extracted. Separately, Hedges' $g$ was calculated between groups after regressing each measure on the two technical covariates alone and retaining the residuals. This residual-based effect size is distinct from the adjusted regression coefficient.

Separately, demographic sensitivity analyses fitted ordinary least squares models of the Core and each principal spectral measure with diagnostic group, age at the analyzed MEG session, and sex as predictors, using HC3 robust standard errors. The demographic and technical adjustments were evaluated separately; neither changed the classification inputs.

\subsubsection{Empirical Sensor-Noise Perturbation}

Eleven participants with noise recordings available for this additional perturbation analysis (five controls and six ADHD) were included to assess noise propagation into source-space measures, without between-group inference. Empirical MEG noise was assembled using contiguous 10-s block bootstrap sampling to match the baseline recording length and added to baseline sensor-space signals at 20, 10, 5, and 0 dB SNR. Five independent noise realizations were generated per participant and SNR. Source reconstruction and feature extraction were repeated for Core, frontal-control Theta, inter-network aperiodic gradient, and superior-parietal aperiodic exponent.

For each measure and SNR, Spearman correlation and mean absolute error (MAE), normalized by the baseline standard deviation across all 42 participants, were calculated relative to baseline and summarized across noise realizations. Theta-band network-role stability was not recomputed in this experiment.

\subsubsection{Beyond-Core Complementarity}

Spearman and group-adjusted partial Spearman correlations described monotonic associations with Core. Separately, the following Core-adjusted model tested whether each expanded measure $k$ retained ADHD-related group information:

\begin{equation}
X_{ik}=\beta_0+\beta_1C_i+\beta_2D_i+\varepsilon_i.
\label{eq:beyond_core}
\end{equation}

Here, $C_i$ is the Core and $D_i$ denotes diagnostic group. HC3 robust standard errors were used to test the group coefficient $\beta_2$ after Core adjustment. Multiple tests were corrected using the Benjamini--Hochberg false discovery rate (FDR) procedure.

Measures showing relatively stable Beyond-Core effects underwent leave-one-subject-out sensitivity analysis. Each participant was excluded in turn, yielding 42 refitted estimates on the original measure scales to assess dependence on influential individuals.


\subsection{Individual-Level ADHD Diagnostic Feasibility}

Core-only classification served as the baseline for an exploratory within-cohort comparison with multidimensional source-space information. Forty repeated stratified splits assigned 28 participants (14 per group) to development and 14 (7 per group) to held-out testing; all comparisons used identical splits.

To assess the contribution of expanded feature inputs under a matched classifier, Core-only and the full seven-measure representation were first compared using the same L2-regularized logistic-regression pipeline with training-only standardization. For each input set, the regularization parameter ($C=0.01, 0.1, 1,$ or $10$) was selected by three-fold out-of-fold ROC-AUC within the development subset, and the threshold (0.35--0.65 in steps of 0.05) was selected from development-set predictions by balanced accuracy. Each fitted classifier was then evaluated on the unchanged held-out subset.

The final multidimensional ensemble considered three subsets of the seven measures in Table~\ref{tab:source_measures}: all seven measures (All7); Core, frontal Theta, aperiodic gradient, and superior-parietal exponent (Spectral4); and these four plus Theta network-role stability (Compact5). L2-regularized logistic regression was used with training-only standardization. The regularization parameter $C$ for All7 was selected within the development subset in the preceding modeling stage, whereas $C=0.3$ was fixed for Spectral4 and Compact5. Each subset's predictions averaged one full-development model and 16 equally weighted models fitted to class-stratified 85\% subsamples.

Fusion weights and thresholds were selected using three repetitions of stratified three-fold out-of-fold predictions within the development subset. Seven predefined weight triplets for (All7, Spectral4, Compact5) were considered: $(1,0,0)$, $(0.75,0.25,0)$, $(0.5,0.5,0)$, $(0.75,0,0.25)$, $(0.5,0,0.5)$, $(0.5,0.25,0.25)$, and $(0.4,0.3,0.3)$. Thresholds ranged from 0.38 to 0.62 in increments of 0.02.

The internal utility combined accuracy, ROC-AUC, and the minimum of sensitivity and specificity with weights of 0.55, 0.35, and 0.10. The reference configuration was All7 with a threshold of 0.5. A candidate was accepted only if its utility improved by at least 0.012, accuracy did not decrease, sensitivity and specificity each decreased by no more than $1/14$, and ROC-AUC decreased by no more than 0.025 relative to this reference. Otherwise, the reference configuration was retained.

Standardization, model fitting, and configuration selection did not use held-out participants. The selected threshold and fusion weights were applied unchanged to testing. ROC-AUC was the principal discrimination metric; accuracy, sensitivity, and specificity were also computed for each held-out subset and summarized across 40 splits as mean $\pm$ SD.

Because the splits reused the same 42 participants and model development followed exploratory comparisons in this cohort, results indicate exploratory within-cohort performance, not independent clinical validation.

\subsection{Statistical Analysis}

Participants were the independent units
in all primary statistical analyses.
Group differences were reported using
means, standard deviations, Cohen's $d$,
and Hedges' $g$, with two-sided
permutation tests.
The primary Core comparison used
10,000 permutations.
Multiple parallel tests were corrected
using the Benjamini--Hochberg FDR procedure.

Spearman rank correlations were used
for continuous variables, with partial
Spearman correlations when adjusting
for diagnostic group.
Core-adjusted and demographic regressions used
HC3 robust standard errors; technical-factor
regressions used conventional OLS inference.
Key Beyond-Core
effects were assessed using
leave-one-subject-out sensitivity analyses.
Matched EEG--MEG experiments used
within-subject paired statistics.
Sensor-noise analyses reported Spearman
correlations and normalized MAE at each SNR.

Unless otherwise specified,
two-sided $p<0.05$ indicated
statistical significance.
Diagnostic performance was summarized
using the mean and SD of 40 repeated
held-out evaluations, with ROC-AUC
as the principal discrimination metric.
Repeated evaluations were not treated
as independent participant-level
observations. All analyses were
performed in Python, primarily using
MNE-Python, NumPy, SciPy, pandas,
scikit-learn, and statsmodels.

%% file: results.tex
\section{Results}

\subsection{MEG Measurement Advantages Enable
Robust Source-Space Network Recovery}

We first assessed whether MEG measurement
properties supported the source-space network
representation used in subsequent analyses.
Under the reference head model, both EEG
and MEG recovered the predefined task-control,
DMN, and Core variations.
Differences emerged primarily under modeling
uncertainty rather than reference conditions.

When skull conductivity deviated from the
reference value, EEG exhibited greater Core
recovery errors than MEG.
Under low-conductivity mismatch, median
normalized MAE was 0.279 for EEG versus
0.065 for MEG; under high-conductivity
mismatch, the corresponding values were
0.419 and 0.074.
MEG showed lower errors in 11 of 12
subject-condition comparisons across six
participants and two mismatch conditions.
Thus, MEG better preserved source-space
network information under conductivity
uncertainty rather than showing universally
superior recovery.

Source-orientation analysis revealed a
complementary difference.
The median direction-selectivity index was
0.686 for MEG versus 0.064 for EEG,
with the same direction of difference in
all six participants (Wilcoxon $p=0.0313$).
MEG showed a substantially lower normalized
radial response (0.187 vs.\ 0.983), whereas
tangential responses remained relatively
strong.

\begin{figure*}[!t]
\centering
\includegraphics[width=\textwidth]{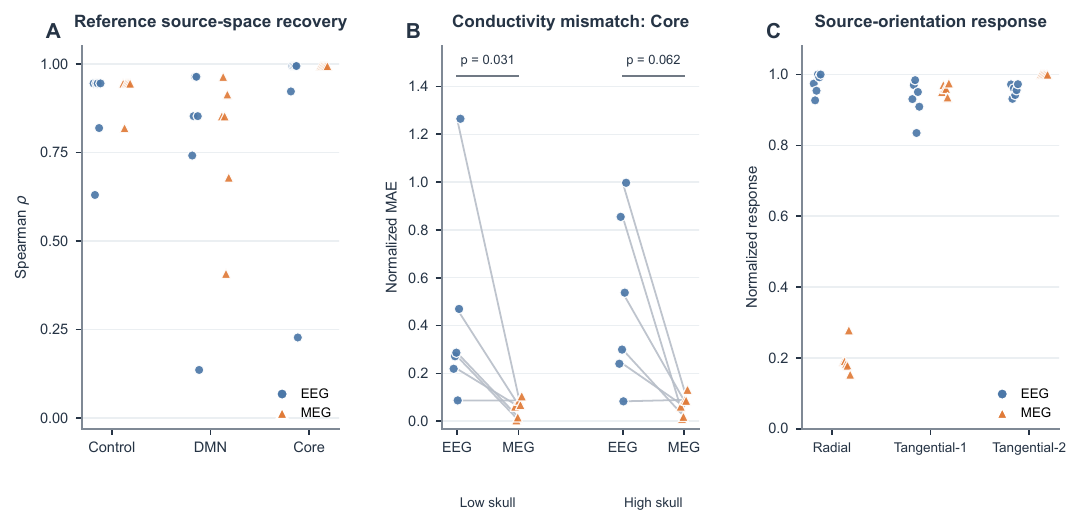}
\caption{MEG measurement properties and
source-space network recovery.
(A) Spearman correlations between predefined
and reconstructed task-control, DMN, and Core
information under the reference head model.
Blue circles and orange triangles denote EEG
and MEG, respectively.
(B) Normalized mean absolute error (nMAE)
of Core recovery under low- and
high-skull-conductivity mismatch.
Each point represents one participant, and
gray lines connect paired EEG--MEG results.
The displayed $p$-values refer to paired
modality comparisons.
(C) Normalized EEG and MEG responses to
radial and two orthogonal tangential
source orientations.
Source locations and strengths were held
constant across orientation conditions.}
\label{fig:meg_properties}
\end{figure*}

Together, these findings indicate that MEG's
relevance to this framework lies in its
lower sensitivity to conductivity mismatch
and source-orientation selectivity.
These properties support subsequent
cross-subject source-space network analysis
(Fig.~\ref{fig:meg_properties}).

\subsection{Core Alpha Imbalance Captures ADHD
Abnormality but Compresses Its Internal Sources}

We next examined the relative Alpha state
between the DMN and task-control-related
systems.
Core was substantially lower in ADHD than
in controls ($-0.7584\pm0.2542$ vs.\
$-0.4511\pm0.3308$), with a large group
effect (Cohen's $d=1.042$, Hedges'
$g=1.022$) and a significant two-sided
permutation test ($p=0.0014$).
Thus, Core captured a pronounced alteration
in the relative Alpha activity of the
two systems.

However, a network difference does not
reveal the contribution of each system.
Decomposition showed markedly lower
task-control-related Alpha in ADHD than
in controls ($-0.5465\pm0.3635$ vs.\
$-0.0694\pm0.3940$; $d=1.259$,
$g=1.235$, $p=0.0004$).
DMN-related Alpha was also numerically
lower in ADHD ($0.2159\pm0.2940$ vs.\
$0.3839\pm0.2858$), but the effect was
smaller and not statistically significant
($d=0.580$, $g=0.569$, $p=0.0638$).
The component-wise results suggest a
more pronounced frontal-control Alpha
contribution than an isolated DMN
Alpha difference.

\begin{figure*}[!t]
\centering
\includegraphics[width=\textwidth]{
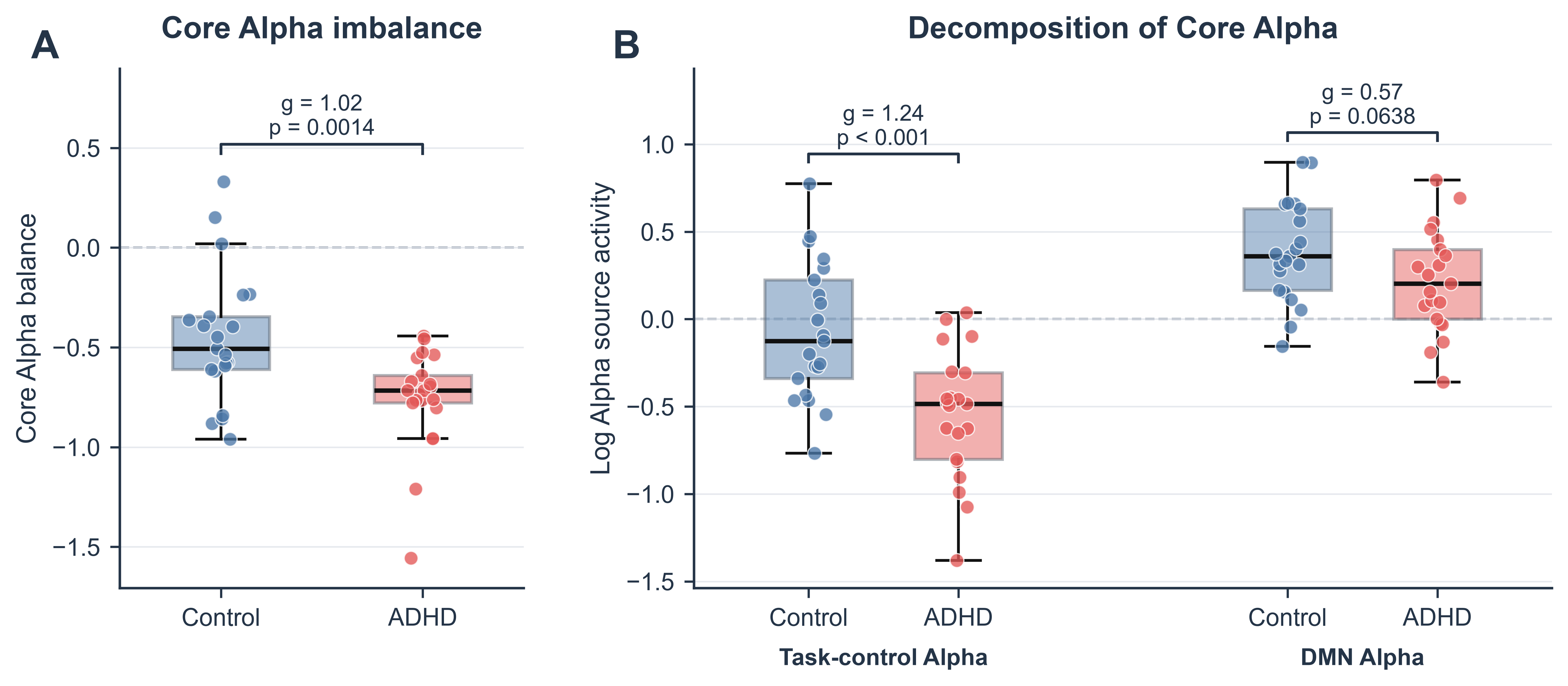}
\caption{Core Alpha imbalance and
decomposition of its internal sources.
(A) Subject-level distribution of the
DMN--task-control Alpha Core in controls
and ADHD participants.
(B) Separate distributions of
task-control-related and DMN-related
log Alpha amplitudes.
Blue and red indicate controls and ADHD,
respectively; overlaid dots represent
individual participants.
Boxes summarize the distributions, with
horizontal black lines indicating medians.
Brackets indicate group comparisons,
annotated with Hedges' $g$ and
two-sided permutation $p$-values.
Task-control-related Alpha showed stronger
group separation than the isolated
DMN Alpha component.}
\label{fig:core_decomposition}
\end{figure*}

Different combinations of task-control
and DMN activity can produce similar
Core values.
Although Core summarizes the dominant
network imbalance, it cannot fully
represent the underlying source-space state.
We therefore examined whether multidimensional
and Beyond-Core analyses could reveal
additional ADHD-related information
(Fig.~\ref{fig:core_decomposition}).

\subsection{Multidimensional Expansion Reveals
Information Beyond the Core}

Beyond spatial decomposition, four measures
extended Core: frontal-control Theta,
inter-network aperiodic gradient,
superior-parietal exponent, and Theta-band
network-role stability. They probe the
Core's DMN--frontal systems or related
parietal cortex through distinct spectral
and temporal-network properties.

All four measures showed ADHD-related
group differences
(Table~\ref{tab:expanded_results}).
Frontal-control Theta was lower in ADHD
than in controls
($-0.532\pm0.234$ vs.\
$-0.200\pm0.277$;
Hedges' $g=1.271$, $p=0.0001$),
as was the inter-network aperiodic gradient
($-0.194\pm0.242$ vs.\
$-0.038\pm0.221$;
$g=0.660$, $p=0.0344$).
The superior-parietal aperiodic exponent
was higher in ADHD
($0.993\pm0.347$ vs.\
$0.793\pm0.217$;
$g=-0.681$, $p=0.0299$),
whereas Theta-band network-role stability
was lower
($0.967\pm0.040$ vs.\
$0.987\pm0.012$;
$g=0.671$, $p=0.0241$).

Diagnostic-group-adjusted partial Spearman
correlations with Core were $\rho=0.280$,
$-0.237$, $0.076$, and $-0.222$
(FDR-adjusted $q=0.206$, 0.206, 0.640,
and 0.206). These estimates describe monotonic
association with Core, distinct from the
Beyond-Core group effects tested below.

Beyond-Core analysis tested whether
these measures retained group information
after Core adjustment.
Three of the four expanded measures
retained significant ADHD-related
group effects: frontal-control Theta
($\beta=-0.710$, HC3 $p=0.0110$,
$q=0.0326$), inter-network aperiodic
gradient ($\beta=-0.904$, $p=0.0163$,
$q=0.0326$), and Theta-band network-role
stability ($\beta=-0.867$, $p=0.0348$,
$q=0.0465$).
The superior-parietal aperiodic exponent
did not reach significance after Core
adjustment ($\beta=0.463$, $p=0.1286$,
$q=0.1286$).
These results indicate differing levels
of statistical support for complementary
ADHD-related information, rather than
establishing significant differences
between the effects of individual measures.

\begin{figure*}[!t]
\centering
\includegraphics[width=0.88\textwidth]{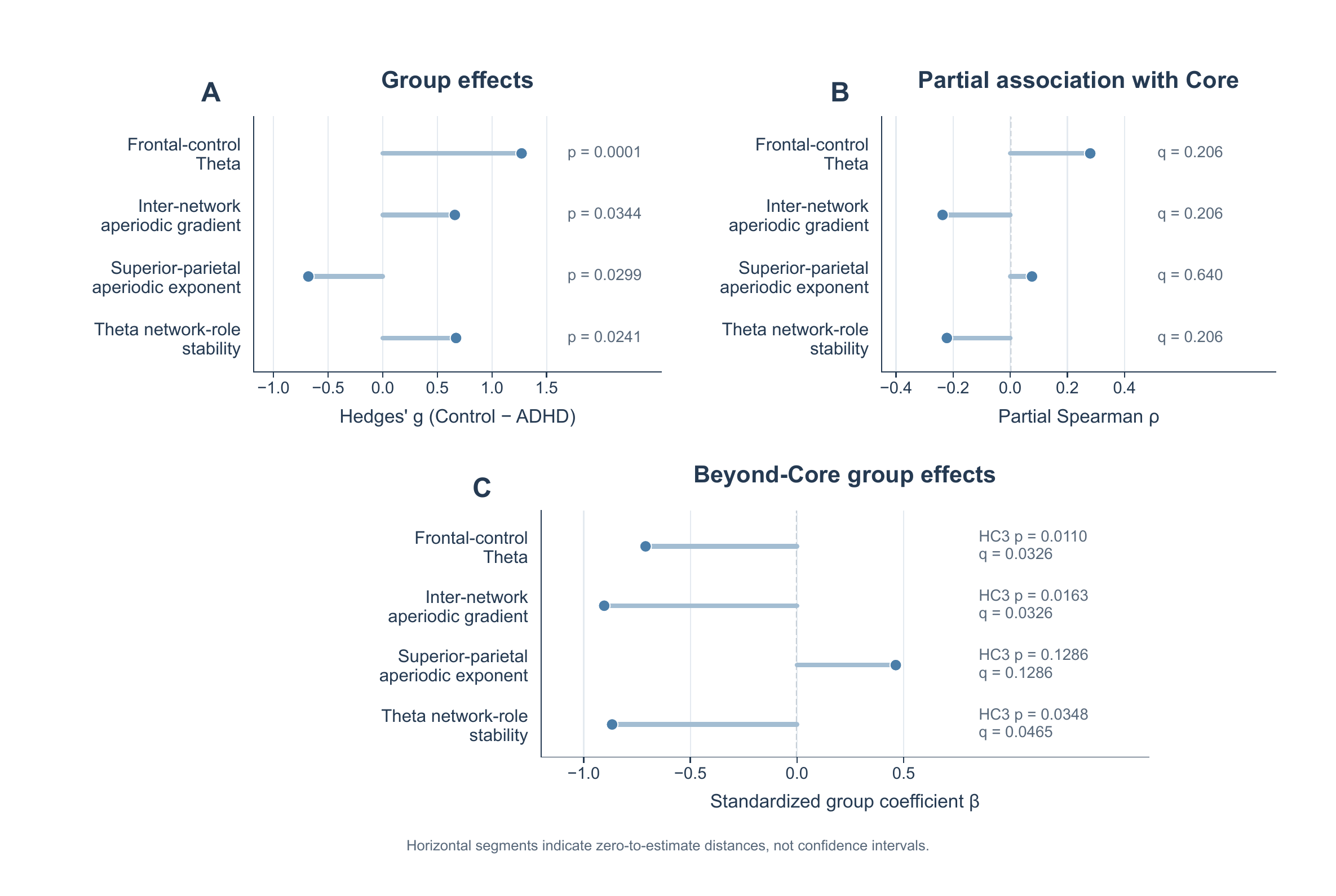}
\caption{Multidimensional expansion and
Beyond-Core information.
(A) ADHD--Control group effects for four
expanded source-space measures, expressed
as Hedges' $g$ (Control minus ADHD),
with permutation $p$-values.
(B) Partial Spearman correlations between
each expanded measure and the Core after
adjustment for diagnostic group.
Displayed $q$-values were corrected using
the Benjamini--Hochberg false discovery
rate (FDR) procedure.
(C) Standardized diagnostic-group
coefficients from Core-adjusted regression
models, annotated with HC3 robust
$p$-values and FDR-adjusted $q$-values.
In panels A--C, horizontal segments extend
from zero to the corresponding estimates
and do not represent confidence intervals.
}
\label{fig:beyond_core}
\end{figure*}

Overall, the expanded measures provided
different degrees of ADHD-related
information beyond the Core.
The Core-adjusted and sensitivity
analyses support multidimensional
characterization of the underlying
neurophysiological state without
assuming that all expanded measures
have equivalent statistical effects
(Fig.~\ref{fig:beyond_core}).

\begin{table*}[!t]
\centering
\caption{Group-Level Results of the
Expanded Source-Space Measures}
\label{tab:expanded_results}
\footnotesize
\renewcommand{\arraystretch}{1.2}
\setlength{\tabcolsep}{4pt}
\begin{tabular}{
p{0.29\textwidth} c c c c
}
\toprule
\textbf{Measure} &
\textbf{Control (Mean $\pm$ SD)} &
\textbf{ADHD (Mean $\pm$ SD)} &
\textbf{Hedges' $g$} &
\textbf{Permutation $p$} \\
\midrule
Frontal-control Theta &
$-0.200\pm0.277$ &
$-0.532\pm0.234$ &
$1.271$ & $0.0001$ \\
Inter-network aperiodic gradient &
$-0.038\pm0.221$ &
$-0.194\pm0.242$ &
$0.660$ & $0.0344$ \\
Superior-parietal aperiodic exponent &
$0.793\pm0.217$ &
$0.993\pm0.347$ &
$-0.681$ & $0.0299$ \\
Theta-band network-role stability &
$0.987\pm0.012$ &
$0.967\pm0.040$ &
$0.671$ & $0.0241$ \\
\bottomrule
\end{tabular}
\end{table*}

\subsection{Reliability and Sensitivity of the Expanded Representation}

\begin{figure*}[!t]
\centering
\includegraphics[width=0.91\textwidth]{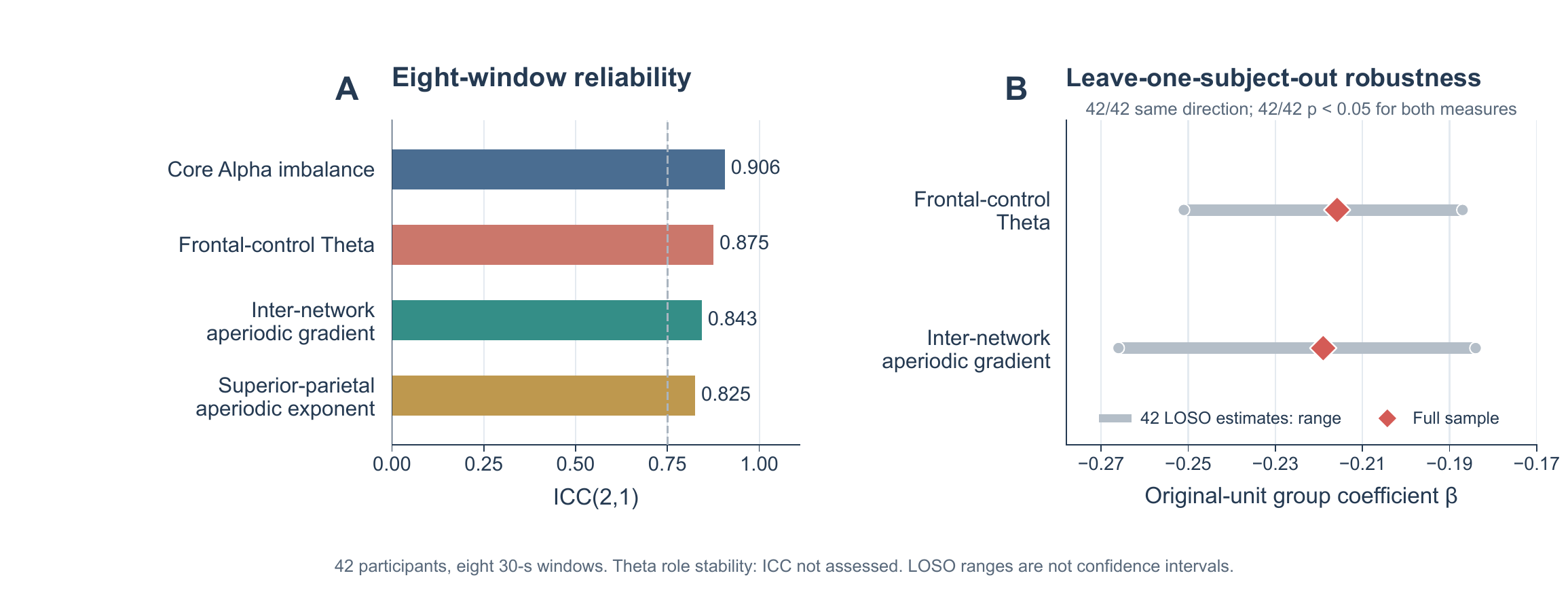}
\caption{Reliability and subject-exclusion sensitivity of source-space MEG measures. (A) ICC(2,1) across eight nonoverlapping 30-s windows for Core Alpha imbalance, frontal-control Theta, inter-network aperiodic gradient, and superior-parietal exponent in 42 participants; the dashed line marks ICC = 0.75. (B) Leave-one-subject-out robustness of the Core-adjusted group associations for frontal-control Theta and inter-network aperiodic gradient. Gray segments indicate the ranges of original-unit group coefficients across 42 exclusions; red diamonds indicate the full-sample estimates. Theta network-role stability was not assessed by window-level ICC.}
\label{fig:reliability_robustness}
\end{figure*}

To assess whether the observed network measures and principal Beyond-Core effects were driven by short recordings or individual observations, we examined window-level reliability and leave-one-subject-out (LOSO) sensitivity (Fig.~\ref{fig:reliability_robustness}).

Across eight nonoverlapping 30-s windows in 42 participants, the Core Alpha imbalance showed high ICC(2,1) reliability (0.906). Frontal-control Theta (0.875), inter-network aperiodic gradient (0.843), and superior-parietal aperiodic exponent (0.825) also exceeded 0.82 (Fig.~\ref{fig:reliability_robustness}A). The superior-parietal reliability estimate was based on window-specific spectral fits. Theta-band network-role stability was not evaluated using window-level ICC because repeated estimates were unavailable.

LOSO analysis further examined whether the Beyond-Core group associations of frontal-control Theta and the aperiodic gradient depended on a single participant. Both effects preserved their direction and statistical significance in all 42 exclusions (Fig.~\ref{fig:reliability_robustness}B); the respective coefficient ranges on the original measure scales were $-0.251$ to $-0.187$ and $-0.266$ to $-0.184$. Together, the window-level and subject-exclusion analyses support the repeatability of the principal measurements and the stability of these Core-adjusted associations.

Additional sensitivity checks considered technical covariates, age and sex, and empirical sensor-noise perturbations. Technical adjustment retained significant group terms for the Core ($\beta=-1.404$, $p=0.0032$) and frontal-control Theta ($\beta=-1.514$, $p=0.0007$); effects for the other expanded measures were more sensitive to technical covariates. After age and sex adjustment, the Core ($p=0.0002$), frontal-control Theta ($p=0.0007$), and aperiodic gradient ($p=0.0300$) remained significant, whereas the superior-parietal exponent did not ($p=0.3212$). Under empirical sensor-noise perturbation, the Core retained a Spearman rank correlation of $\rho=0.845$ even at 0 dB; noise sensitivity varied across other measures. These checks indicate that the strongest results do not depend solely on a single participant or short-window fluctuation, while their robustness is measure-dependent.

\subsection{Exploratory Within-Cohort ADHD Classification}

We assessed the individual-level relevance of the multidimensional source-space representation through exploratory ADHD classification. Core-only and multidimensional approaches were compared using 40 identical stratified held-out splits, each comprising 28 development and 14 test participants. All within-split model fitting and configuration selection used development data only.

In the matched logistic-regression comparison, adding the six expanded measures to Core increased mean ROC-AUC from 0.7607 to 0.8995 ($+0.1388$) and accuracy from 71.43\% to 77.50\%, using the same splits and tuning procedure. This comparison supports a discriminative contribution from multidimensional inputs without relying on ensemble fusion.

The final multidimensional approach achieved a mean ROC-AUC of $0.9051\pm0.0598$ versus $0.7607\pm0.1046$ for Core-only (Table~\ref{tab:diagnosis}). Higher AUC occurred in 36 of 40 splits, with two ties and two decreases. Mean accuracy increased from $71.43\pm10.85\%$ to $79.82\pm9.13\%$, sensitivity from $74.64\pm13.91\%$ to $80.00\pm17.36\%$, and specificity from $68.21\pm22.53\%$ to $79.64\pm13.29\%$. These repeated within-cohort results indicate exploratory individual-level discrimination, not independent external validation.

\begin{table*}[!t]
\centering
\caption{Individual-Level ADHD Identification Performance}
\label{tab:diagnosis}
\footnotesize
\renewcommand{\arraystretch}{1.2}
\setlength{\tabcolsep}{6pt}

\begin{tabular}{
p{0.20\textwidth}
p{0.30\textwidth}
p{0.34\textwidth}
}
\toprule
\textbf{Metric} &
\textbf{Core-only} &
\textbf{Multidimensional Representation} \\
\midrule
ROC-AUC &
$0.7607\pm0.1046$ &
$0.9051\pm0.0598$ \\
Accuracy (\%) &
$71.43\pm10.85$ &
$79.82\pm9.13$ \\
Sensitivity (\%) &
$74.64\pm13.91$ &
$80.00\pm17.36$ \\
Specificity (\%) &
$68.21\pm22.53$ &
$79.64\pm13.29$ \\
\bottomrule
\end{tabular}

\vspace{3pt}
\parbox{0.87\textwidth}{
\footnotesize
Values are mean $\pm$ SD across 40 repeated stratified held-out evaluations, each with 28 development and 14 test participants. Models used identical partitions. Repeated test subsets were drawn from the same 42-participant cohort and did not constitute independent external validation.
}
\end{table*}

%% file: discussion.tex
\section{Discussion}

This study addressed information obscured when DMN--task-control activity is reduced to a single Alpha imbalance value. Spatial decomposition resolved the Core's component contributions, while physiologically related spectral and temporal measures characterized properties not represented by the Alpha contrast. A matched logistic-regression comparison improved mean ROC-AUC from 0.7607 to 0.8995 with expanded inputs, and the final multidimensional approach reached 0.9051. The main contribution is an interpretable link between low-dimensional network imbalance, targeted multidimensional expansion, and exploratory individual-level discrimination.

\subsection{From MEG Measurement Properties
to Source-Space Network Representation}

We first examined the physical measurement
properties supporting MEG source-space analysis.
In matched EEG--MEG simulations, MEG showed
lower network-representation errors under
skull-conductivity perturbations and greater
source-orientation selectivity, consistent
with previous studies~\cite{ref23,ref24,ref25,ref26}.
These properties are relevant because both
Core and its expanded measures depend on
cross-regional and cross-subject source-space
comparisons, where forward-model uncertainty
may propagate through inverse reconstruction
into the resulting network measures.

Both modalities recovered network information under
the reference model. MEG's relative robustness
to conductivity mismatch and its source-orientation
selectivity instead provide measurement-level
support for the representation, without implying
universally superior source localization or
ADHD classification performance.

\subsection{Neurophysiological Basis and
Information Expansion of the Core Imbalance}

The principal neurophysiological finding
was altered DMN--task-control Alpha balance
in ADHD.
Previous fMRI and electrophysiological studies
have reported abnormalities in the DMN,
task-related networks, and their
interactions~\cite{ref05,ref06,ref07,ref08,
ref09,ref10,ref11,ref12}.
However, network imbalance alone cannot
identify the contribution of each system.
Our decomposition showed a stronger
task-control-related Alpha alteration,
whereas the isolated DMN Alpha difference
was weaker.
The Core abnormality should therefore
not be interpreted simply as increased
DMN activity; changes in task-control-related
activity appear to contribute substantially
to the relative network state.

Alpha oscillations are associated with
information gating, functional inhibition,
and network-state regulation~\cite{ref30},
but their amplitude does not directly
measure a single cognitive function.
Accordingly, we interpret the Core
as altered relative electrophysiological
organization between DMN and task-control
systems rather than attributing it
to one region or psychological process.

Nevertheless, the Core compresses two Alpha
components into a single difference: individuals
with similar Core values can have distinct frontal-
control and DMN contributions. Spatial decomposition
exposes this otherwise obscured information. We then
extended the representation to spectral and temporal
properties that the Alpha contrast does not measure.
Frontal-control Theta captured additional oscillatory
information consistent with its association with
cognitive control~\cite{ref31}.

Spectral expansion complemented band-limited
oscillatory measures with empirical spectral-slope
descriptors, which may capture distinct aspects
of neural activity~\cite{ref28,ref29}.
Our aperiodic findings indicate that
similar Alpha network states may coexist
with different background spectral
organization. The inter-network gradient
examines spectral differences between the
two systems comprising the Core, while the
superior-parietal exponent extends the
analysis to attention-related cortex outside
that contrast. These measures were selected
to broaden physiological coverage, not to
assume that each provides independent
ADHD-related information.
Thus, spectral expansion adds physiological
descriptors that the Alpha contrast
does not directly encode, rather than
emphasizing any single measure.
Importantly, the aperiodic exponent
is not a direct measure of excitation--
inhibition balance, and our findings
do not establish specific changes
in that balance.

Theta-band network-role stability characterized
temporal consistency of network participation,
beyond information available from static spectra.
Core-adjusted results showed selective rather than
uniform complementarity, supporting a multidimensional
representation without treating every expanded measure
as an equally informative biomarker.

\subsection{Reliability, Information Gain, and Individual-Level Implications}

The Core and three principal expanded measures showed high short-window reliability (ICC $>0.82$). The Core-adjusted effects of frontal-control Theta and the inter-network aperiodic gradient also persisted across all 42 leave-one-subject-out exclusions, reducing concern that these findings arose from one influential participant. Technical and demographic adjustments provided further support for the Core and frontal Theta associations, although robustness varied among the other descriptors. Together, these checks reinforce the principal expanded representation without assuming uniform robustness across all measures.

Crucially, the matched logistic-regression comparison increased mean ROC-AUC from 0.7607 to 0.8995 and accuracy from 71.43\% to 77.50\% when expanded features were added. The additional gain to 0.9051 ROC-AUC and 79.82\% accuracy with the final ensemble was comparatively modest. Thus, improved discrimination was evident before ensemble fusion, supporting the value of multidimensional physiological inputs rather than attributing the result solely to modeling complexity. These repeated within-cohort evaluations remain exploratory and require independent validation for clinical generalization.

\subsection{Limitations and Future Work}

The cohort was relatively small, with potential age
imbalance and limited clinical phenotyping.
Demographic and technical adjustments supported
the principal Core and frontal Theta findings,
although sensitivity varied across expanded
measures. Anatomical ROIs approximated functional
networks, and medication and comorbidity data
were not consistently available. Because the
classification comparisons reused one exploratory
cohort, independent validation of the locked
representation and modeling pipeline is needed.

%% file: conclusion.tex
\section{Conclusion}

A single DMN--task-control Alpha imbalance
measure cannot fully characterize its underlying
source-space neurophysiological state.
Using MEG, we decomposed the core imbalance
measure (Core) into its spatial contributions
and extracted complementary information
along spectral and temporal-network dimensions.
Our findings showed that task-control-related
regions exhibited more pronounced ADHD-related
differences in Alpha activity than the DMN.
After adjusting for the Core, Theta activity
in task-control-related regions and the
inter-network empirical aperiodic gradient
still showed significant ADHD-related group
differences, revealing neurophysiological
information not fully captured by the
single imbalance measure.
In matched logistic-regression comparisons,
multidimensional inputs increased mean ROC-AUC
from 0.7607 to 0.8995; the final ensemble
achieved a mean ROC-AUC of 0.9051 and a
mean accuracy of 79.82\%.
Overall, MEG-based multidimensional source-space
expansion not only provides further insight
into the neurophysiological composition of
ADHD-related network imbalance but also
improves exploratory individual-level
discrimination.
Independent cohorts are still needed
to establish generalizability.